\documentclass{aa}
\usepackage{graphicx}
\usepackage{natbib}
\usepackage{txfonts}
\begin{document}
   \title{An S-shaped Arc in the Galaxy Cluster RX J0054.0-2823\thanks{Based 
		on observations obtained at the European Southern
		Observatory, La Silla, Chile and ESO archive files}}

   \author{C. Faure
          \inst{1, 4}
          \and
          E. Giraud
          \inst{2}
          \and
           J. Melnick
          \inst{3}
          \and
           H. Quintana
          \inst{4}          
          \and
           F. Selman
          \inst{3}
          \and
           J. Wambsganss
          \inst{1}}

   \institute{Zentrum f\"ur Astronomie der Universit\"at Heidelberg (ZAH), M\"onchhofstr. 12-14, 69120 Heidelberg, Germany
              \and
              Laboratoire Physique Th\'eorique et Astroparticules,
              UMR5207 In2p3/Montpellier University, F-34095 Montpellier
                           \and
              European Southern Observatory, Alonso de C\'ordova 3107,
              Santiago, Chile
                           \and
	      Department of Astronomy and Astrophysics, P. Universidad 
              Catolica de Chile, Casilla 306, Santiago, Chile
	      }

   \date{accepted 29/09/2006}
\authorrunning{C. Faure et al.}
\titlerunning{An S-shaped arc in RX J0054.0-2823}
\abstract{The center of the galaxy cluster RX J0054.0-2823 at z = 0.292 is a 
   dynamically active region which includes an interacting system 
   of three galaxies surrounded by a large halo of intra-cluster light.
  We report here the discovery of an S-shaped feature of total length 11\arcsec\, in the central region of this cluster and discuss its physical nature. We test the gravitational lensing assumption by doing a mass modelling of the central part of the galaxy cluster.  We very naturally reproduce position and form of this S-shape feature as a gravitationally lensed background object at redshift    between 0.5 and 1.0. We conclude that the lensing nature is the very probable explanation for this S-shaped arc; the ultimate proof will be the spectroscopic confirmation by measuring the high redshift of this elongated feature with surface brightness V$\sim$24~mag~arcsec$^{-2}$.} 


 \maketitle


\section{Introduction}

Giant luminous arcs were discovered in 1986/87 
(Lynds \& Petrosian 1986, Soucail et al. 1987) 
and found to be highly distorted background galaxies, 
magnified and often multiply imaged by the gravitational
lensing action of an intervening galaxy cluster (Paczy\'nski 1987). Meanwhile, more than 100 giant arcs systems
have been found 
(Gladders, Yee \& Ellingson 2002 and references therein)  and studied both 
to infer the cluster mass distribution and 
-- using the magnification effect -- 
to investigate the high redshift galaxy population 
(e.g., Colley, Tyson \&  Turner 1996).
The ensemble of arcs was also used for statistical inferences
regarding consistency with various cosmological models
(e.g., Bartelmann et al. 1998; Golse, Kneib \& Soucail 2002; Soucail, Kneib \& Golse 2004;
Wambsganss, Bode  \& Ostriker 2004;
Sand et al. 2005). 
 
Contrary to the implication of the name, 
the shapes of highly distorted background galaxies 
are not always curved like an arc: lots
of radial arcs have been found, elongated
features extended in the radial direction; 
a straight arc was discovered early on as well (Wambsganss
et al. 1989).
Here we report the discovery of a 
strangely shaped arc candidate.  
Our modelling shows that a very natural explanation
for this S-shaped feature is a background galaxy
gravitationally lensed by the moderately  massive 
X-ray cluster RX J0054.0-2823, with the shape particularly
affected by the  three prominent central galaxies,
and the cluster ellipticity.\\


\section{Observations and data reduction}

The galaxy cluster 
RX J0054.0-2823
was identified as part of the
160 square degree ROSAT serendipity survey
(Vikhlinin et al. 1998, Mullis et al. 2003).
Our imaging observations of  RX J0054.0-2823 
consist of a series of 24 V-band and 39 I-band 
images, with an exposure time of 300 s each, 
acquired during two photometric dark nights on 2000 
September 26 and 27, using SUSI2 at the ESO NTT. 
The total exposure time was 18900s.
SUSI2 covers a field of $5' \times 5'$ with two CCDs at a 
pixel scale of $\rm 0.0805'' pixel^{-1}$. 
The CCDs were binned $2 \times 2$. In order to correct for fringing, sky 
background and CCD variations, the observations were carried out with an 
alternating dithering pattern as described in Melnick, Selman \& Quintana (1999). 
In this method, 
the cluster center is placed on one of the CCDs, say CCD1, while
CCD2 simultaneously measures the sky background.  
Then the cluster is centered on CCD2, and CCD1 
is used to measure the sky. The exposure is thus divided in a large number of 
integrations, nodding consecutively on CCD1 and CCD2, 
and off-setting the telescope 
by small random amounts to dither the images. 
The differences between consecutive images allow to correct for fringing, to built and subtract for darks 
and to remove for most systematic effects. 
The frames are flat-fielded by using a master flat.
The final images, which are the medians of individual 300s exposures,  
are calibrated using Landolt stars (1992).  
A maximum variation by a factor 
of 2 was found in the I-band sky background over the two nights, with 3 
main groups of values at 19.2, 19.5 and $\rm 19.9~mag~arcsec^{-2}$. 
The background noise 
depends on the number of individual images at a given position: in the central 
overlap area  the 1$\sigma$ isophotes are at
$V = 27.0$ and $I = 25.7~ \rm mag~arcsec^{-2}$ respectively.
The connection between pixels of real objects allows clear detections 
at these magnitudes.  The full field size of our final image is 
$9.5' \times 5.5'$.  

\begin{figure}
   \includegraphics[width=8.5cm]{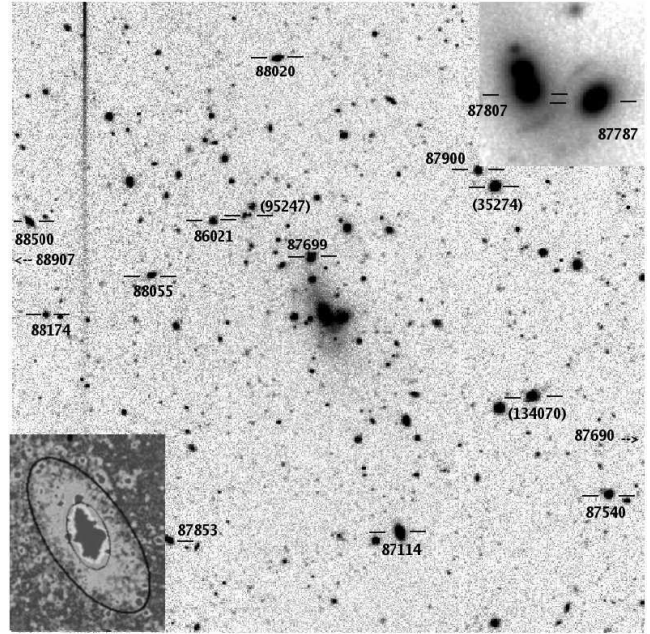}
      \caption{The V-band image (4\arcsec$\times$4\arcsec) in log intensity scale 
		centered on RX J0054.0-2823.  The total exposure time  is 7200s, 
		and the seeing 0.8\arcsec\, FWHM. The radial 
		velocities are super-imposed to the galaxies (in $\rm km~sec^{-1}$). 
		Velocities of galaxies outside the field are marked by arrow. The velocity 95247~km~s$^{-1}$ corresponds to two galaxies. Therefore we have measured the velocity dispersion for 18 objects in the field.
		The upper insert is the central 16\arcsec$\times$16\arcsec region  
		where the S-shaped feature  is visible between the three central galaxies.  	
		The lower insert is the 59\arcsec$\times$75\arcsec central region
		binned 9$\times$9 pixels and it shows the intergalactic light. 
		The inner (outer) ellipse delineates the 
		$V=\rm 27 (31) mag~ arcsec^{-2}$-contour.  
		Both have similar ellipticities  $\epsilon=1-b/a\sim$0.4 and orientations, 
		$\theta\sim$20$^{\circ}$ ($\sim$30$^{\circ}$). North is to the top, East is to the left.}
         \label{FigSarc}
   \end{figure}

The galaxy spectra of RX J0054.0-2823 which we retrieved from 
the ESO archives (program 64.O-0455(A); P.I. H. Quintana) 
were originally obtained using EFOSC 
in multi-object spectroscopic mode (MOS) at the ESO 3.6m telescope. 
Two consecutive MOS spectra of the same galaxies with 1800s exposure time each, 
had been acquired during the night 1999 November 8 
with atmospheric seeing  of $0.9''$ and $0.8''$, respectively.
They had been obtained with slitlets of $1.5''$ and Grism~\#6
which provides a wavelength coverage of
380-850 nm at a resolution of 2 nm (R $\sim 300$). 
The detector was binned 
$2 \times 2$ at readout giving a spectral dispersion
of 0.4 nm $\rm pixel^{-1}$ and a spatial scale of $0.31''~\rm pixel^{-1}$. 
The data were 
analyzed using the  context LONG within MIDAS. The
two-dimensional spectra were corrected for bias and flat-field and wavelength 
calibrated using HeAr lines. 
One-dimensional spectra were then extracted 
after sky subtraction. Redshifts were measured by fitting the continuum and 
the identified lines by using the MIDAS context ALICE. They 
are based on [OII], the CaII H and K doublet, the Mgb 
band, NaID, and the Balmer lines. 
Eighteen individual galaxy redshifts were thus obtained 
(all indicated in Fig.~\ref{FigSarc}) out of which
14 turned out to be cluster members (see also redshift histogram in Fig.~\ref{FigHisto}). 

\section{The galaxy cluster: images and  redshifts }

\begin{figure}
   \includegraphics[width=8cm]{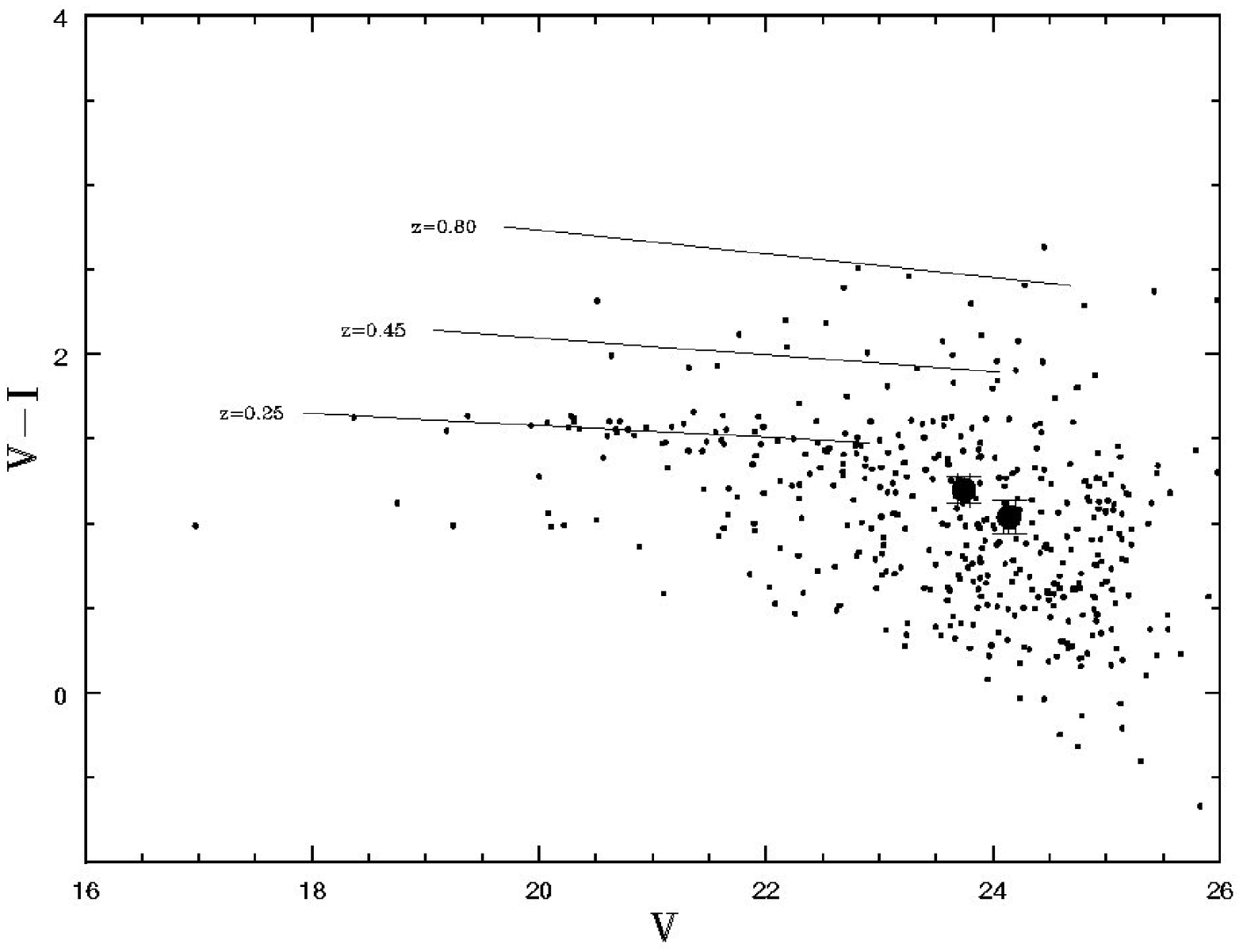}
 \includegraphics[width=8.5cm]{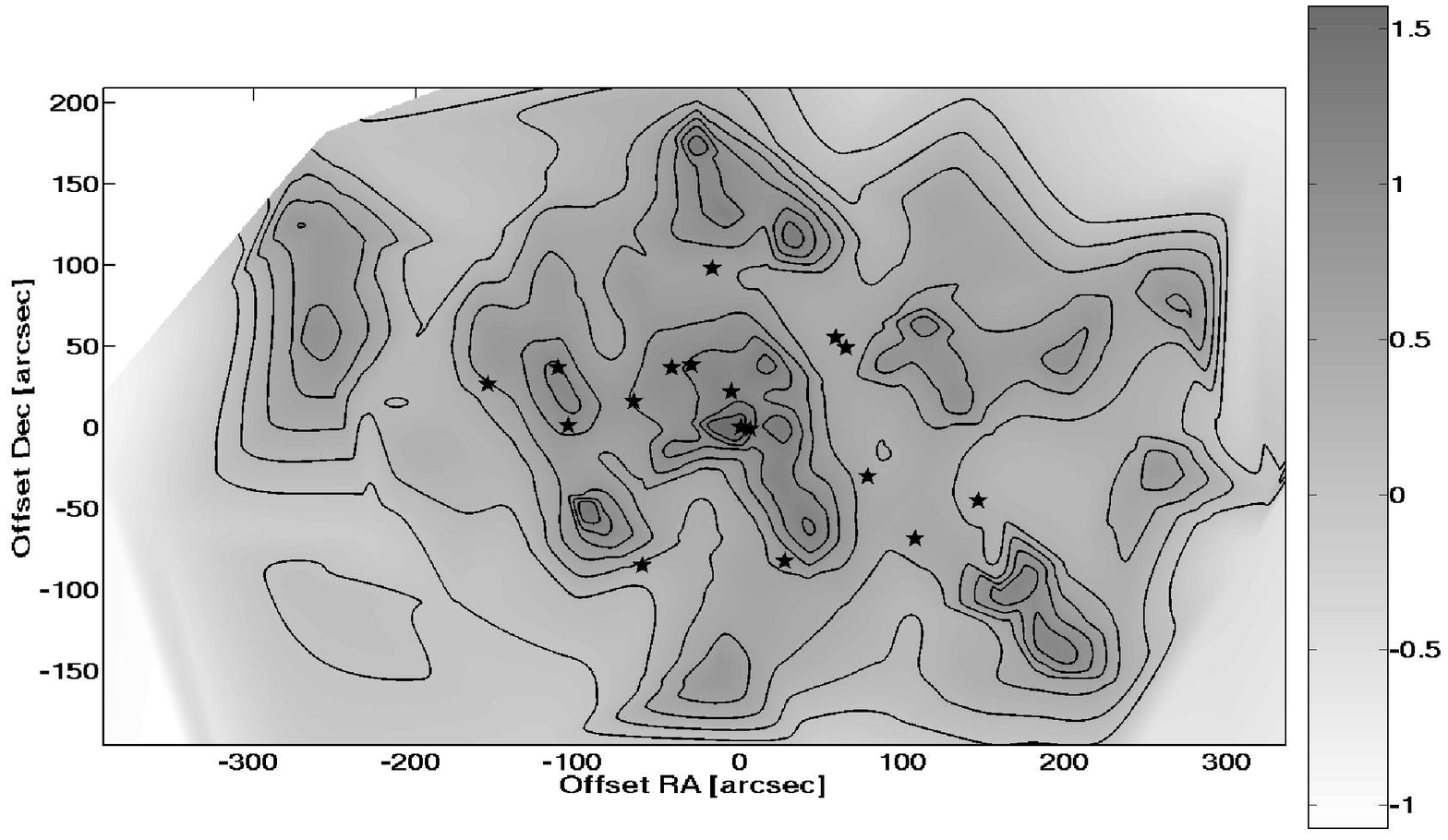}
      \caption{\label{redseq}
	Top panel: Cluster color-magnitude diagram showing the positions of the red-sequence, the 
	two arcs (big black dots), and the locus of red-sequence expected for redshifts of
	0.25, 0.45, and 0.80 according to Kodama et al.(1998).
	Bottom panel: Contour map of the $\rm log_{10}$ of galaxy density selecting only
	galaxies within 0.25~mag from the red-sequence. The stars
	mark the position of galaxies with measured redshift. The density is in unit of galaxies per square arcminute, and the contour are separated by 0.2~dex intervals. North is to the top, East is to the left.}
   \end{figure}

The V-band image of galaxy cluster RX J0054.0-2823 is  shown in 
Fig.~\ref{FigSarc}. On a 4\arcmin\, (1.3~h$_{75}^{-1}$~Mpc)  scale, the 
main feature is an apparently elongated distribution of galaxy 
traced by a few bright galaxies at PA$\sim$30$^{\circ}$ 
with a concentration around 3 major objects in close contact, 
surrounded by a common envelope, 
which is also the location of the X-ray emission
detected by ROSAT (Mullis et al. 2003).  The central galaxy triplet (see top right insert of Fig. \ref{FigSarc}) is 
made of a close pair of elliptical galaxies (labeled \#131 and \#139 
in Fig. \ref{modelfig}),
and a more elongated object (\#128).
On a scale of 25\arcsec\, ($\sim$ 0.14$\rm{h_{75}^{-1}~Mpc}$),
our deep images show that the central objects are surrounded by extended intra-cluster light elongated  in the direction of the distribution of the galaxies
described above (cf. Fig.~\ref{FigSarc}, bottom left  insert).
The size of this intra-cluster light component  at the V =27~mag~arcsec$^{-2}$ isophote is
$\sim$30\arcsec$\times$16.7\arcsec, and 2 to 3 times
larger at the V=31~mag~arcsec$^{-2}$ isophote. 
There are several very
faint clumps in this region surrounded by the diffuse intra-cluster light.
 The ROSAT X-ray map is too noisy to measure an 
elongation but its core 
radius is about r$_{core}$=37$\pm$6\arcsec (or  200$\pm$32~h$_{75}^{-1}$kpc)  from Vikhlinin et al.
(1998).
With redshifts for only 14 galaxies and with a limited
spatial coverage it is difficult to estimate the mass of
the cluster. We use the standard virial mass 
estimator \cite{limber1960,carlberg1996}, but we complement
our estimates with an independent check from the measured
X-ray luminosity for this cluster:

\begin{displaymath}
M_\mathrm{v} = {3\over G} \sigma_{1D}^2 r_\mathrm{v},
\end{displaymath}\noindent where $\sigma_{1D}$ is the line-of-sight
velocity dispersion, and $r_\mathrm{v}$ is the virial radius
defined as $r_\mathrm{v}=\pi R_h/2$, with $R_h$ the projected
harmonic radius
\begin{displaymath}
R_h = \left< {1\over \left|\left|\mathbf{R}_i - \mathbf{R}_j\right|\right| } \right>_\mathrm{Pairs},
\end{displaymath}\noindent where $\mathbf{R}_i$ is the projected galaxy position vector. 
The galaxies used to perform this analysis are chosen to be at a distance equal or lower to 0.25~mag  to the red-sequence at redshift z=0.25. This selection  criteria is  illustrated in Fig. \ref{redseq}. 

The above estimates are valid only when the distribution
of galaxies is spherically symmetric, and in our case we
note a considerable elongation in  the density distribution.
We do only an order or magnitude estimate in this section,
pointing out that it has been noticed in the past that
spherical dynamic models can reproduce accurately the
dynamics of \emph{slightly} elongated systems (Sand, Treu \& Ellis 2002; Kronowitter et al. 2000).

\begin{figure}
   \includegraphics[width=8.5cm]{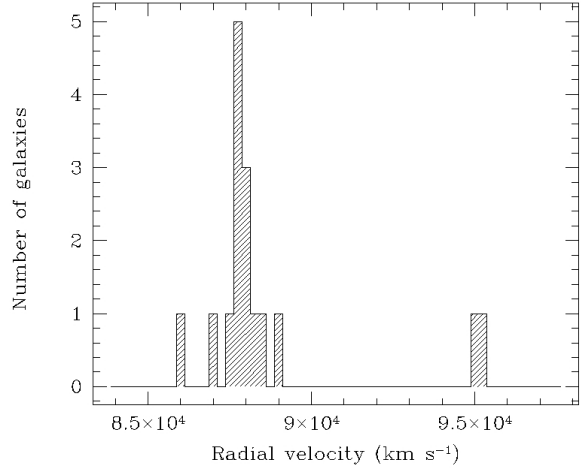}
      \caption{Velocity histogram of the 16 galaxies in the central region
	 of RX J0054.0-2823 (bin size 250 km s$^{-1}$; two galaxies outside range). 
	 }
         \label{FigHisto}
   \end{figure}

The redshift distribution of the 18 objects is shown in Fig. \ref{FigHisto}:
14 of the 18 measured galaxies are cluster members with velocities around $\sim$88000km~s$^{-1}$. The remaining four  galaxies
are a very close pair of faint objects at $\sim$95250~km~s$^{-1}$, 
a background galaxy at 
134070~km~s$^{-1}$ and a foreground galaxy at 35275~km~s$^{-1}$
(not plotted in the histogram). For the cluster members the 
distribution  is well peaked at cz=87775~km~s$^{-1}$
with a dispersion $\sigma$=675~km~s$^{-1}$. 
The average  coincides with the mean of  two of the main central galaxies (\#128 and \#131), in good 
agreement with the redshift z=0.292 given by Mullis et al. 
(2003) for the cluster. 
The internal error in the galaxy redshifts is $\Delta$v=209$\pm$30~km~s$^{-1}$, so 
the intrinsic line-of-sight (rest frame) radial velocity dispersion of the cluster is
$\sigma_{1D}=\sqrt{\sigma^2-\Delta{\rm v}^2}$/(1+z)=497~km~s$^{-1}$, and
the 3D dispersion $\sigma_{3D}=\sqrt{3}~\sigma_{1D}$=860~km~s$^{-1}$.
The harmonic radii of the galaxies with velocities and of the galaxies
in the red-sequence are 410~$h_{75}^{-1}$~kpc$^{-1}$ and 710~$h_{75}^{-1}$~kpc$^{-1}$,
respectively. With this value for the velocity dispersion, and with 
the above values for the harmonic radius we obtain a virial cluster mass 
in the range $\mathrm{(1-2)\times10^{14}~h_{75}^{-1}~ M_\odot}$.
The luminosity of the galaxies in the red sequence within the
corresponding radii are 
$\mathrm{1.76\times10^{11}~h_{75}^{-2}~ L_{Vodot}\ and\ 3.06\times10^{11}~h_{75}^{-2}~ L_{V\odot}}$.
The derived M/L ratios are $\mathrm{570~h_{75}~ M_\odot/L_{V\odot}}$,
and $\mathrm{650~h_{75}~ M_\odot/L_{V\odot}}$, respectively.  This large M/L values have to be seen as upper limit as they are derived considering only galaxy part of the red-sequence.

The X-ray luminosity of the cluster is $L_X=0.63 \times 10^{44}~\rm~h_{75}^{-2}~ erg~ s^{-1}$ 
(Vikhlinin et al. 1998). According to
Reiprich \& B\"{o}ringer (2002), with this luminosity and the mass determined above,
RX~J0054.0-2823 is right in the correlation between X-ray luminosity
and cluster virial mass in the region of the least massive clusters.

The approximate linear arrangement of the galaxies, and the clumpy
general appearance suggest that the region is probably  
dynamically young and that  the triple central galaxy system could be the 
merging progenitor of a future single large cD. In addition to the X-ray emitting 
hot gas, the central potential well appears to be filled with smooth intra-cluster 
stellar light and the faint sub-galaxian clumps could be 
material ejected from the interacting system. 
A composite spectrum in the wavelength range $365-420$nm
was built with the 14 cluster galaxy spectra in order to characterize the
presence of star formation following the method of Dressler et al. 2004.
This spectrum indicates that there
has been no significant star formation in this region for several Gyr
(e.g. weak H$\delta$ and [OII]).


\section{The S-shaped arc}

As shown in Fig.~\ref{FigSarc} (insert top right) and in Fig.~\ref{modelfig} (top left panel) there is 
a giant arc  with
a total length of $\sim$11\arcsec and an unresolved width ($\leq$ 0.70\arcsec) superimposed on the 
intra-cluster light. The arc-like feature as a shape of an ''S'', slightly inclined South-East to North-West. It embraces the South and East part of galaxy \#131 and the West and North part of galaxy \#128. The bright northern and southern extensions of the arc have a surface brightness of V$\sim$24~mag~arcsec$^{-2}$, while there is a drop in luminosity in the region that links them.  \\ 
      
 The (V-I)-color of the arc is $\sim1.1\pm0.1$~for
the North component and  $\sim1.0\pm0.1$ for the
fainter South component (see Fig.~\ref{redseq}, top panel).  
There are various possibilities  for the physical origin of this S-shape arc. 
The color of the arc is consistent
with  being late type gravitationally lensed galaxies at high redshift. But it is  also
consistent with being cluster member(s) in the faint end of the
luminosity function being tidally deformed in an interaction
with the central multiple system.
In that case, this  S-shaped feature could be either tidal debris from the merging of the 
massive central galaxies, or one or two tidally disrupted cluster members.
At V$\sim$24~mag~arcsec$^{-2}$ it would be much brighter
than the tidal features found in Virgo (e.g. Mihos et al. 2005), 
specially if we consider the Tolman dimming which at this redshift
should be approximately 1~mag. Thus, although possible we consider 
the tidal debris origin to be unlikely. 
A tidally disrupted galaxy interacting with the bright central galaxies
is a possibility that cannot be dismissed. A disc galaxy could
in principle create a long tidal tail in the interaction with
one of these galaxies which could then be distorted by the other.
Although this scenario is highly unlikely a priori, we must
remember that it was precisely the peculiar geometry of the system that
caught our attention in the first place.
In what follows we explore the gravitational lens scenario by
modelling the arc-shape with the known galaxy and cluster potentials.

\subsection{Gravitational lens modelling} \label{modelling}

In order to model the S-shape arc as one or more  gravitationally lensed background galaxies  we need to know  the gravitational potentials 
of the putative deflectors. 
These can be estimated using the structural parameters of the galaxies 
as given by their photometry and velocity dispersions. 
The latter have been estimated from the H and K line-profile widths corrected for instrumental
broadening by means of the (weak) emission lines. 
The velocity dispersions turn out to be similar
$\rm\sim370~kms^{-1}$ for the two main galaxies \#128 and \#131, corresponding to 
masses $M_{24}=(3.5-3.8)\times\rm10^{11}~h_{75}^{-1}~M_\odot$ and
M/L ratios $(M/L)_{24}\approx22~\rm h_{75}^{-1}~(M/L)_\odot$, measured at  
V=24~mag~arcsec$^{-2}$.
No spectra were taken for galaxy \#139, hence no velocity dispersion is measured.

 \begin{figure}
   \includegraphics[width=8.5cm]{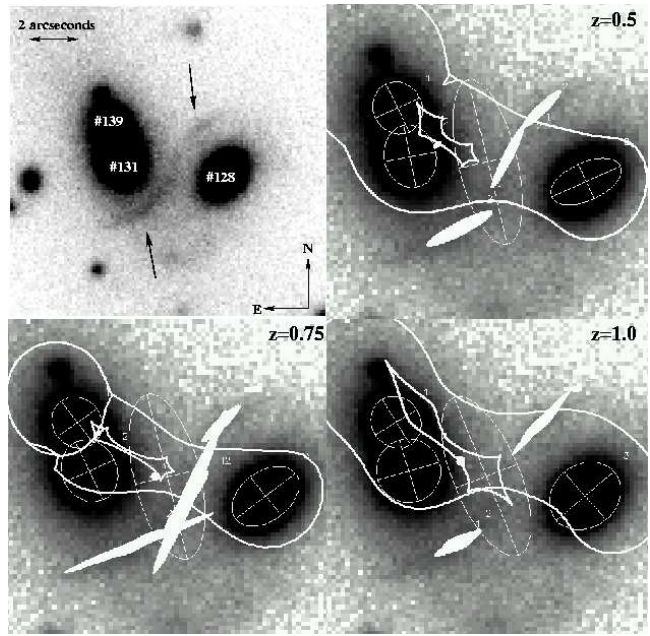}
  \caption{\label{modelfig} Top left panel: The S-shape feature in the 10\arcsec$\times$10\arcsec\, V-band image of the center of RX J0054.0-2823. The names of the galaxies are labeled as in the text and in Table \ref{model}. The arrows point at the bright regions of the arc.  Top right panel and bottom panels: the lens model for three different source redshifts 
	(indicated in the upper-right corner of each image)
superimposed on the central 6\arcsec$\times$6\arcsec\, V-band image.  North is to the top, East is to the left. 
The ellipsoids filled in white show the  location of the images (the source position is represented by a white dot) for the best lens mass model.  
The white ellipsoids indicate the 
location, ellipticity, and orientation of the lensing 
galaxies (\#128, \#131 and \#139) and of the 
galaxy cluster as given in Table~\ref{model}. For the galaxy cluster the size is reduced to optimize its visibility in the figure. The white bold solid lines indicates the inner and outer caustic lines for the best lens mass model.}
 \end{figure}

We want to test whether the S-shaped arc can be the distorted, potentially 
multiple, image of one or more background galaxies. 
Our simplified lens potential consists of four components at z = 0.292:
the cluster, and the three central galaxies: \#128, \#131 and \#139 (see Fig.~\ref{modelfig}). 
These components are fitted by Pseudo-Isothermal Mass Distributions
(PIEMD; Kassiola \& Kovner 1993, Kneib et al. 1996) 
defined by position, core radius, cut radius, orientation, 
ellipticity, and velocity dispersion.
Since the source redshift is not known,
we have computed models for four different source redshifts: 
z$_s$ = 0.5, 0.75, 1.0 and 2.0. \\

Since the arc has two bright regions (see Figs. \ref{modelfig}, top left panel), 
we initially  assumed that it is built by the gravitationally distorted 
images of two background objects.
With these assumptions we used the LensTool package (Kneib et al. 1993)  
to model the arc in  
an iterative way varying the input parameters within the ranges allowed by 
the observations. For the cluster potential the only initial constraint was 
the velocity dispersion,  but since the models are relatively insensitive 
(within reasonable values) to the cluster center position, we chose to 
keep it fixed at a value close to the centroid of the three main 
galaxies (cf. Fig.~\ref{modelfig} and Table \ref{model}).  
We iterated  until we found solutions that satisfied the following criteria: 
(a) input parameters consistent with observations; 
(b) lensed images match the morphology of S-arc.
The code naturally converges to a single source for the two arc components, therefore we supposed that the arc is formed by the distorted images of a single background source.\\

For the fit of the best lens potential, we allow the 3D-velocity dispersion to vary in a 20\% error-bar range around the value 870~km~s$^{-1}$ for the galaxy cluster and  370~km~s$^{-1}$ for galaxies \#128 and \#131. Galaxy \#139 being fainter than the two other galaxies of the model, we assume that its mass is lower and  put an upper limit on its velocity dispersion of 300~km~s$^{-1}$.

\subsection{Results}

\begin{table}
\renewcommand{\arraystretch}{0.9}
\centering
\caption{\label{model} Parameters of the gravitational lens model:
Column 1: Redshift of the source. Column 2: Lens component.
Column 3 and 4:  Lens position relative to the brightest part of the northern arc component.
Column 5: Lens ellipticity.
Column 6: Lens position angle.
Column 7:  Lens velocity dispersion.
}

\begin{tabular}{ l l c c c c c c c c }
Source   & Lens & $\Delta$x(\arcsec) & $\Delta$y(\arcsec) 
	& $\epsilon$&$\theta$($^\circ$) & $\sigma_v$~(km~s$^{-1}$) 
	 \\
\hline
\hline
 z$_s$=0.5&\#128   & -0.81 & -1.37  & 0.49 & -65   & 350    \\
          &\#131   & 2.43  & -0.89  & 0.20 & 11    & 401   \\
          &\#139   & 2.66  &  0.02  & 0.10 & 30    & 252  \\
          &Cluster & 1.0   & -1.0   & 0.66 & 13    & 940   \\
\hline
z$_s$=0.75 &\#128   & -0.81 & -1.37   & 0.34 & -51 & 354   \\
           &\#131   & 2.43  & -0.89   & 0.16 & 36  & 360  \\
           &\#139   & 2.66  &  0.02   & 0.10 & 25  & 217  \\
           &Cluster & 1.0  & -1.0     & 0.62 & 13  & 811 \\
\hline
z$_s$=1.0 &\#128   &  -0.81 & -1.37    & 0.10 & -43 & 230    \\
          &\#131   &  2.43  & -0.89    & 0.02 & 30 & 150 \\
          &\#139   &  2.66  &  0.02    & 0.10 & 34 & 285  \\
          &Cluster &  1.0  & -1.0      & 0.51 & 28 & 618    \\

\hline
\hline

\end{tabular}
\end{table}

Our best-fit models for a gravitational lensing scenario are shown in Fig.~\ref{modelfig} for three values of the 
source redshift (z$_s$ = 0.5, 0.75, 1.0).
  A large sample of good fit can be found for the observed parameter error-bars. This illustrates the robustness of the gravitational lensing hypothesis and  the high probability of having such a S-shaped arc formed in the heart of such mass potential configuration.   



The parameters of our best-fitting models are  tabulated in Table~ \ref{model}: 
for each source redshift we report the 
position relative to the center of the  North bright component of the arc  ($\Delta x$, $\Delta y$), 
	 the ellipticity (defined as $\epsilon$=1-$b$/$a$, where $a$ is the semi-major axis and $b$ is the semi-minor axis), 
	the position angle ($\theta$,  positive North to East counter-clockwise) and 
	 and the velocity dispersion ($\sigma_v$) 
	of the lens components are given.

Our best fits are obtained for source redshifts 
0.5 $\le$ z$_s <$  1.0.  
For z=1, we match the galaxy and cluster PA and ellipticity only for velocity dispersions much lower than the measured one (see Table \ref{model}). 
Moreover, we were not able to obtain good fits for $z_{s} \ge 1$.
In fact, the model parameters of the lens components 
depart increasingly from the observations  as the source 
redshift increases. 

According to model for source redshift a z$_s$=0.5 and 0.75, the arc would be  built by 3 images of a single background galaxy: two images in opposite orientation (mirror images) would built the northern part of the arc, one single image  would be the southern part. The luminosity drop observed in the northern  part of the arc would reflect a decreasing surface brightness region in the galaxy source, that would also be, in the lens plane, the location where the two northern images are partially overlap.  

%

We conclude that gravitational lensing is a natural and very 
likely explanation for the S-shaped arc 
observed in RX J0054.0-2823.


\section{Summary and conclusion}

We report the discovery of an S-shaped arc in the very center 
of the galaxy cluster RX J0054.0-2823.
We discuss various possibilities for the physical origin of the feature.
Modelling the potential formed by three massive central 
galaxies plus the cluster itself,  
we obtained very good fits to location and shape of this arc for 
a gravitationally lensed galaxy source at 0.5 $\le$z$_s<$1.0.  
These models show that the S-shaped feature is likely to be the triple gravitational image 
of a single background galaxy.  
There is not a unique satisfying  solution for the modelling. This implies that lensing is a very natural explanation for the S-arc feature for the  configuration observed for the galaxies and the galaxy cluster. 

All the  best-fitting potential of the cluster have
an ellipticity and orientation  consistent with 
the distribution of both the diffuse intra-cluster light, 
and of the X-ray emission.
This, together with the elongated galaxy distribution, and the 
relatively large mass inferred from the Virial theorem, suggests that 
the cluster is dynamically young, and probably the result of 
a recent merging.
 
Given the success of the lensing modelling, the alternative explanations of
tidally induced compression of the stellar intra-cluster medium, or a highly distorted cluster spiral appear rather unlikely. 
The gravitational lensing explanation  only works if the redshift of the source 
is lower than $z_{s}\sim 1$.  Therefore, a straightforward test  for the lensing hypothesis is to measure the redshift of the arc.

\begin{acknowledgements}
C.F. and J.W. are supported by 
the European Community's Sixth Framework 
Marie Curie Research Training Network Programme, 
Contract No. MRTN-CT-2004-505183 ``ANGLES'' and 
granted by the ECOS-CONYCIT Commitee.  
H.Q. is grateful for partial support from 
the FONDAP Astrophysics Center.
\end{acknowledgements}

\end{document}